\documentclass[11pt]{article}
\usepackage{cite}
\usepackage{amsmath,amsfonts,amssymb}
\pdfoutput=1
\usepackage[small,bf,hang]{caption}
\usepackage{slashed}
\input epsf.sty
\usepackage{epsfig}
\usepackage[titletoc,toc]{appendix}
\usepackage{xcolor}

\def\hybrid{
        \topmargin -20pt
        \oddsidemargin 0pt
        \headheight 0pt \headsep 0pt
        \textwidth 6.55in 
        \textheight 9.5in 
        \marginparwidth .875in
        \parskip 5pt plus 1pt \jot = 1.5ex}

\hybrid

 \csname
@addtoreset\endcsname{equation}{section}

\def\moth{\mathsurround=0pt}
\newdimen\zo \zo=0pt

\def\tick{\leaders\hrule height 0.5ex depth 0pt \hskip 0.5pt}
\def\upboxfill{$\moth \setbox\zo\hbox{\tick}%
  \hskip 3pt\hbox to 0pt{$\tick$\hss}\hrulefill \hbox to 7.5pt{$\tick$\hss}$}

\def\dtick{\leaders\hrule height .34pt depth 0.5ex \hskip 0.5pt}
\def\downboxfill{$\moth \setbox\zo\hbox{\dtick}%
  \hskip 2pt\hbox to 0pt{$\dtick$\hss}\hrulefill \hbox to 2pt{$\dtick$\hss}$}

\def\id{{\mathbb I}}

\def\bec{\begin{center}}
\def\ec{\end{center}}

\def\be{\begin{equation}}
\def\ee{\end{equation}}
\def\bea{\begin{eqnarray}}
\def\eea{\end{eqnarray}}
\def\ba{\begin{array}}
\def\ea{\end{array}}

\usepackage{hyperref}

\usepackage{tikz}
\usepackage{tikz-cd}
\usepackage{mathabx}

\begin{document}

\begin{titlepage}
	
	\rightline{\tt MIT-CTP-5712}
	\hfill \today
	\begin{center}
		\vskip 0.5cm
		
		{\Large \bf {$A_\infty$ perspective to Sen's formalism}
		}
		
		\vskip 0.5cm
		
		\vskip 1.0cm
		{\large {Atakan Hilmi F{\i}rat}}
		
		{\em  \hskip -.1truecm
			Center for Theoretical Physics \\
			Massachusetts Institute of Technology\\
			Cambridge, MA 02139, USA\\
			\tt \href{firat@mit.edu}{firat@mit.edu} \vskip 5pt }

		\vskip 1.5cm
		{\bf Abstract}
		
	\end{center}
	
	\vskip 0.5cm
	
	\noindent
	\begin{narrower}
		\noindent
		Sen's formalism is a mechanism for eliminating constraints on the dynamical fields that are imposed independently from equations of motion by employing spurious free fields. In this note a cyclic homotopy associative algebra underlying Sen's formalism is developed. The novelty lies in the construction of a symplectic form and cyclic $A_\infty$ maps on an extended algebra that combines the dynamical and spurious fields. This algebraic presentation makes the gauge invariance of theories using Sen's formulation manifest.
	\end{narrower}
	
\end{titlepage}

\tableofcontents

\section{Introduction}

Eliminating restrictions on fields as much as possible is almost always desirable in field theories. Recently Ashoke Sen has introduced a novel mechanism for eliminating constraints that are imposed on top of equations of motion (such as the self-duality constraints for $p$-form fields) motivated by string field theory (SFT)~\cite{Sen:2015hha,Sen:2015uaa,Sen:2015nph,Sen:2019qit}, for reviews refer to~\cite{new_review,Zwiebach:1992ie,deLacroix:2017lif,Erler:2019loq,Erbin:2021smf,Maccaferri:2023vns}. Sen's formalism for lifting constraints introduces spurious fields to the theory that couple to the interacting fields at the level of the kinetic term. The relevant equations of motion remain the same while imposing the constraint separately is no longer required, but one has to pay the price of having free fields that decouple from the spectrum entirely. This approach has been implemented in various theories so far; ranging from eliminating the self-duality constraint for the $5$-form field strength of IIB supergravity~\cite{Sen:2015nph,Chakrabarti:2022jcb,Chakrabarti:2023czz}, to the chiral boson in $2d$\cite{Chakrabarti:2020dhv,Chakrabarti:2022lnn}, to the construction of the Ramond sectors in super SFTs~\cite{Sen:2015hha,Sen:2015uaa}, to eliminating the notorious level-matching condition of closed SFT~\cite{Okawa:2022mos,Erbin:2022cyb}. 

In this brief note we construct a cyclic homotopy (associative) algebra, or $A_\infty$ algebra, underlying Sen's formalism and its relation to the cyclic homotopy algebra whose elements are constrained. The latter algebra is reviewed in section~\ref{sec:constraint}. The approach here eventually boils down to introducing a new symplectic form and explicit $A_\infty$ maps on an unconstrained algebra that contains not just the original field $\Psi$, but also Sen's spurious field $\widetilde{\Psi}$, see section~\ref{sec:noconstraint}. This algebraic framework makes the gauge invariance of Sen's formalism manifest and its connection to the homotopy algebras and Batalin-Vilkovisky (BV) formalism becomes more apparent.  Although we exclusively work with $A_\infty$ algebras as a proof of principle, we expect that the generalization to the remaining species of homotopy (loop) algebras is a straightforward exercise.

In particular, we present our results in the coalgebra language~\cite{Vosmera:2020jyw}, which would help to make the computations for the theories that employ the Sen's trick more efficient and streamlined. For instance, satisfying the (quantum) master equation of the BV formalism and the (loop) homotopy algebra relations can be argued to be equivalent~\cite{Maccaferri:2023gcg}. Therefore having a (loop) homotopy algebra for the Sen's formalism would directly imply its BV quantization and this would eliminate the necessity of directly working with the fields like in~\cite{Sen:2015hha}. In a future publication~\cite{KS} we are going to use our coalgebraic techniques for the Sen's formalism to demonstrate that the Kodaira-Spencer theory without the level-matching solves the quantum master equation. This was our main motivation behind this work.

\noindent \textbf{Note added:} As this work is finalized the author has been informed that there are some parallels between the homotopy algebra in section 2.2 of~\cite{Erler:2017pgf} and the construction sketched here. However, contrast to this previous work, our homotopy algebra can be adapted to apply Sen's formalism generically and we present our results in the language of coalgebra for the reasons explained above. These two points constitute the novelty of our work.

\section{Cyclic $A_\infty$ algebra with a constraint} \label{sec:constraint}

We begin the note by reviewing the basics of cyclic homotopy algebras and its relation to the field theory with constraints to facilitate the discussion. The form of the homotopy algebra in this section is motivated from the structures appearing in SFT: the Ramond (R) sector of open superstrings in particular~\cite{Kunitomo:2015usa,Erler:2016ybs}. 

It is highly important to stress that the structure developed here is applicable to~\emph{any} theory whose fields are constrained.\footnote{This may possibly require to discuss $L_\infty$ formulation instead of $A_\infty$ one, but this constitutes a trivial extension as we mentioned in the introduction.} We often give the example of the R sector of open superstrings to provide a context to our abstract construction, but one shouldn't get the impression that the construction is only applicable to them. For this reason, we omit discussing the Neveu-Schwarz (NS) sector and keep its interactions implicit below---as they are not particularly relevant to the constraint and we don't want to complicate our discussion. It is certainly possible to include them with some amount of work however, see~\cite{Erler:2017pgf} for example.

So start by imagining a $\mathbb{Z}$-graded vector space $\mathcal{H}$ over the super-complex numbers $\mathbb{C}^{1 | 1}$ for which the states $| \Psi \rangle = \Psi \in \mathcal{H}$ satisfy the constraint
\begin{align} \label{eq:C}
	XY | \Psi \rangle = | \Psi \rangle \, ,
\end{align}
and the operators $X,Y$ obey
\begin{align} \label{eq:Xyx}
	XYX = X \, , \quad \quad YXY =Y \, .
\end{align}
It immediately follows that the operator $XY$ is a projector $(XY)^2 = XY $. The operators $X,Y$ are fermionic (i.e. degree-even), but $X (Y)$ is assumed to increase (decrease) the grade by $1$. We denote the subspace of $\mathcal{H}$ obeying the constraint~\eqref{eq:C} by $\widehat{\mathcal{H}} = XY \mathcal{H}$. 

These definitions are abstracted from the structures appearing within open super SFT~\cite{Kunitomo:2015usa,Erler:2016ybs}. The grade for this special case is the picture number (plus 1/2) and the degree refers to the Grassmannality (plus 1), while the operators $X,Y$ are given by
\begin{align} \label{eq:XY}
	X = -\delta(\beta_0) \, G_0 + \delta'(\beta_0) \, b_0 \, , \quad \quad
	Y = - c_0 \, \delta' (\gamma_0) \, ,
\end{align}
in terms of the modes of the $bc\beta\gamma$ ghost system and the zero mode of the supercurrent $G_0$. The vector space $\mathcal{H}$ can be taken to be the small Hilbert space.\footnote{Acting the operators $X,Y$ to an arbitrary graded state may be ill-defined like in superstrings. However we ignore this subtlety to keep our discussion general. The actions, equations of motion, and gauge symmetries below still remain well-defined in any case.}. We emphasize our considerations are not just limited to open superstrings: an analogous structure can be found within the level-matched closed bosonic SFT. In that case the grade is the (2 minus) ghost number and $X=b_0^- \delta(L_0^-)$ and $Y= c_0^-$ for instance~\cite{Erbin:2022cyb,Okawa:2022mos}. Here the mode $L_0$ is the zeroth Virasoro charge.

Now suppose that a cyclic $A_\infty$ algebra on the space $\widehat{\mathcal{H}}$ is given. The most efficient way to present this structure is through~\emph{the tensor coalgebra}~\cite{Vosmera:2020jyw,Erbin:2020eyc}
\begin{align}
	T \widehat{\mathcal{H}} \equiv \bigoplus_{n=0}^\infty \widehat{\mathcal{H}}^{\otimes n}
	= \mathbb{C}^{1|1} \oplus \widehat{\mathcal{H}} \oplus 
	(\widehat{\mathcal{H}}  \otimes \widehat{\mathcal{H}} ) + \cdots \, ,
\end{align}
with the natural projection maps
\begin{align}
	\widehat{\pi_n} : T \widehat{\mathcal{H}} \to \widehat{\mathcal{H}}^{\otimes n} \, .
\end{align}

Let us introduce some cooperations on $T \widehat{\mathcal{H}}$ before proceeding. The first one is~\emph{the coproduct} $\Delta: T \widehat{\mathcal{H}} \to T \widehat{\mathcal{H}} \boxtimes T \widehat{\mathcal{H}}$ defined by
\begin{align}
	\Delta (\widehat{\Psi}_1 \otimes \cdots \otimes \widehat{\Psi}_n) 
	\equiv \sum_{m=0}^n (\widehat{\Psi}_1 \otimes \cdots \otimes \widehat{\Psi}_m) \boxtimes
	(\widehat{\Psi}_{m+1} \otimes \cdots \otimes \widehat{\Psi}_n) \, ,
\end{align}
for $\widehat{\Psi}_i \in \widehat{\mathcal{H}}$. This is a linear operation and $\boxtimes$ arises from the splitting provided by the coproduct $\Delta$, which is taken to be distinct from the ordinary tensor product $\otimes$. The coproduct $\Delta$ is~\emph{coassociative}, i.e. it commutes in the diagram
\[
\begin{tikzcd}
	T \widehat{\mathcal{H}} \arrow[rr, "\Delta"] \arrow[dd, "\Delta"]                             &  & T \widehat{\mathcal{H}} \boxtimes T \widehat{\mathcal{H}} \arrow[dd, "\Delta \, \boxtimes \, \boldsymbol{1}"] \\
	&  &                                                                                     \\
	T \widehat{\mathcal{H}} \boxtimes T \widehat{\mathcal{H}} \arrow[rr, "\boldsymbol{1} \, \boxtimes \, \Delta"] &  & T \widehat{\mathcal{H}} \boxtimes T \widehat{\mathcal{H}} \boxtimes  T \widehat{\mathcal{H}}          
\end{tikzcd}
\vspace{1em}
\]
Here $\boldsymbol{1} $ is the identity operator on the tensor coalgebra. Using the coproduct we can also define~\emph{a coderivation} $\boldsymbol{D} :  T \widehat{\mathcal{H}} \to T \widehat{\mathcal{H}} $, which is a linear map satisfying~\emph{the co-Leibniz rule}
\[
\begin{tikzcd}
	T \widehat{\mathcal{H}} \arrow[rr, "\boldsymbol{D}"] \arrow[dd, "\Delta"]                                                                        &  & T \widehat{\mathcal{H}}\arrow[dd, "\Delta"]    \\
	&  &                                       \\
	T \widehat{\mathcal{H}} \boxtimes T \widehat{\mathcal{H}} \arrow[rr, "\boldsymbol{1} \, \boxtimes \, \boldsymbol{D} + \boldsymbol{D} \, \boxtimes \, \boldsymbol{1}"] &  & T \widehat{\mathcal{H}} \boxtimes T \widehat{\mathcal{H}}
\end{tikzcd}
\vspace{1em}
\]
We note that the commutator of two coderivations itself is a coderivation.\footnote{The commutators are always taken to be graded antisymmetric.} Finally, define~\emph{a cohomomorphism} $\boldsymbol{F} :  T \widehat{\mathcal{H}} \to T \widehat{\mathcal{H}} $, which is a linear map that commutes in the diagram
\[
\begin{tikzcd}
	T \widehat{\mathcal{H}} \arrow[rr, "\boldsymbol{F}"] \arrow[dd, "\Delta"]                             &  & T \widehat{\mathcal{H}} \arrow[dd, "\Delta"] \\
	&  &                                                                                     \\
	T \widehat{\mathcal{H}} \boxtimes T \widehat{\mathcal{H}} \arrow[rr, "\boldsymbol{F} \, \boxtimes \, \boldsymbol{F}"] &  & T \widehat{\mathcal{H}}\boxtimes T \widehat{\mathcal{H}}            
\end{tikzcd}
\vspace{1em}
\]
If a cohomomorphism is invertible it is \emph{a coisomorphism}. The similar definitions apply to other tensor coalgebras that appear in this note.

The $A_\infty$ algebra on $\widehat{\mathcal{H}}$ is compactly encoded in an odd nilpotent coderivation in $T\widehat{\mathcal{H}}$
\begin{align} \label{eq:m0}
	\boldsymbol{m}^2 = \boldsymbol{0} \, ,
\end{align}
where $\boldsymbol{0} $ is the zero element of the tensor coalgebra. The coderivation $\boldsymbol{m}$ can be decomposed as
\begin{align} \label{eq:m}
	\boldsymbol{m} = \boldsymbol{m_1} +\boldsymbol{ \widehat{\delta m}} 
	= \boldsymbol{m_1} + \boldsymbol{\widehat{ m_2}} + \cdots  
	= \boldsymbol{m_1} + \sum_{n=2}^\infty \, \boldsymbol{\widehat{m_n}}  \, , \
\end{align}
where the coderivation $\boldsymbol{m_1}$ is associated with the free theory and the rest comes from the $n$-point interactions of the theory, i.e. $\boldsymbol{\widehat{m_n}}\widehat{\pi_m} = 0$ for $m < n$. The coderivation $\boldsymbol{m_1}$ is nilpotent by itself due to being associated with a free theory, $\boldsymbol{m_1}^2 = \boldsymbol{0} $, so
\begin{align} \label{eq:A1}
	\boldsymbol{m_1} \, \boldsymbol{ \widehat{\delta m}} +
	\boldsymbol{\widehat{\delta m}} \, \boldsymbol{m_1} +
	\boldsymbol{\widehat{\delta m}}^2 = \boldsymbol{0}  \, .
\end{align}
We define the multi-linear, degree-odd maps $\widehat{m_n}$ by $\widehat{m_n} = \widehat{\pi_1} \boldsymbol{\widehat{m}} \widehat{\pi_n}$. This collection of maps forms the $A_\infty$ algebra after suspending the degree by one. The map $m_1$ is the BRST operator $Q_B$ in the context of SFT. We take $m_1$ to be defined on~$\mathcal{H}$ to have a notion of a free theory on this space.

We further demand the map $m_1$ to satisfy~\cite{Kunitomo:2015usa,Erler:2016ybs}\footnote{Take note that the operator $Y$ doesn't commute with $m_1$. In the open superstring context $Y=-c_0 \, \delta'(\gamma_0)$ for the Ramond sector~\cite{Kunitomo:2015usa,Erler:2016ybs}. This $Y$ is different from the zero mode of the inverse picture changing operator $\mathcal{Y} = c \, \delta'(\gamma) = c \, \partial \xi \, e^{-2\phi}$, which commutes with the BRST operator $m_1 = Q_B$~\cite{Witten:1986qs,Sen:2021tpp}. The author thanks Ted Erler for explaining this subtle fact. Some discussion along these lines is also provided in~\cite{Alexandrov:2022mmy}.}
\begin{align} \label{eq:comm}
	[ m_1, X ] = 0 \, , 
\end{align}
so that it can be restricted to $\widehat{\mathcal{H}}$, i.e. $m_1$ maps $| \Psi \rangle \in \widehat{\mathcal{H}}$ to an element in $ \widehat{\mathcal{H}}$
\begin{align}\label{eq:arg}
	m_1 | \Psi \rangle &= 	m_1 X Y | \Psi \rangle = X m_1 Y | \Psi \rangle 
	= X Y X m_1 Y | \Psi \rangle  = X Y m_1 X Y  | \Psi \rangle  =  XY m_1 | \Psi \rangle \nonumber \\
	&\implies m_1 | \Psi \rangle \in \widehat{\mathcal{H}} \, ,
\end{align}
using~\eqref{eq:C} and~\eqref{eq:Xyx}. After defining the cohomomorphisms
\begin{align} \label{eq:Coho}
	\boldsymbol{X} \equiv \sum_{n=0}^\infty X^{\otimes n} = 1 + X + X \otimes X + \cdots
	\, , \quad 
	\boldsymbol{Y} \equiv \sum_{n=0}^\infty Y^{\otimes n}  = 1  + Y \otimes Y + \cdots \, ,
\end{align}
the identities~\eqref{eq:Xyx} and~\eqref{eq:comm} can alternatively be expressed at the level of the tensor coalgebra $T \mathcal{H}$
\begin{align} \label{eq:like}
	\boldsymbol{X} \boldsymbol{Y} \boldsymbol{X}  = \boldsymbol{X}  \, , \quad \quad
	\boldsymbol{Y} \boldsymbol{X} \boldsymbol{Y}  = \boldsymbol{Y}  \, , \quad \quad
	[ \boldsymbol{m_1},  \boldsymbol{X} ] = 0 \, .
\end{align}

The $A_\infty$ algebra has to endow a 2-form to provide an action for the theory. We assume that there is one in $\mathcal{H}$ 
\begin{align} \label{eq:OGsymp}
	\langle \omega | : \mathcal{H}^{\otimes 2} \to \mathbb{C}^{1|1} \, ,
\end{align}
which is provided by the BPZ product in the context of SFT for example. It is further assumed to have the properties
\begin{subequations}
\begin{align}
	&\langle \omega | \, \Psi_1 \otimes \Psi_2 = - (-1)^{\Psi_1 \Psi_2} 
	\langle \omega | \, \Psi_2 \otimes \Psi_1 \, , \\
	&\forall \, \Psi_1 \in \mathcal{H} \quad \quad
	\langle \omega |  \, \Psi_1 \otimes  \Psi_2 = 0
	\implies  \Psi_2 = 0 \, , \label{eq:nD} \, \\
	& \langle \omega | \, \widehat{\pi_2} \, \boldsymbol{m_1} 
	= \langle \omega | \, (m_1 \otimes \id + \id \otimes m_1 )=
	0 \, , \label{eq:nD1}
\end{align}
\end{subequations}
for $\Psi_i \in \mathcal{H}$, where $(-1)^{\Psi_i}$ is the degree of $\Psi_i$.  The operators $X$, $Y$ are taken to be anti-cyclic
\begin{align} \label{eq:AntiCyc}
	\langle \omega | \left( X \otimes \id - \id \otimes X \right) =
	\langle \omega | \left( Y \otimes \id - \id \otimes Y \right) = 0 \, ,
\end{align}
which, again, is motivated by the constructions within SFT. Here $\id$ is the identity operation on $\mathcal{H}$. We sometimes use the notation $\omega \, (\Psi_1, \Psi_2) = \langle \omega| \, \Psi_1 \otimes \Psi_2$ for the symplectic forms.

In the context of the constrained SFT the form $\langle \omega |$ requires insertions of $Y$ in order to saturate the zero modes demanded by various kinds of worldsheet anomalies and obtain a nonvanishing quadratic target spacetime action. The relevant anomaly for the superstrings is the picture number anomaly while it is the ghost number anomaly for the level-matching of closed bosonic strings. This means that the following symplectic form has to be used instead
\begin{align} \label{eq:what}
	\langle \widehat{\omega} | : \widehat{\mathcal{H}}^{\otimes 2}  \to \mathbb{C}^{1|1}  \, , \quad \quad
	\langle \widehat{\omega} |
	\equiv \langle \omega|  \, \id \otimes Y
	 = \langle \omega| \, Y \otimes \id \, .
\end{align}

This form is graded antisymmetric
\begin{align} \label{eq:AS}
	\langle \widehat{\omega} | \, \widehat{\Psi}_1 \otimes \widehat{\Psi}_2 &
	= \langle \omega| \, \widehat{\Psi}_1 \otimes Y \widehat{\Psi}_2 
	= - (-1)^{\widehat{\Psi}_1 \widehat{\Psi}_2 }\langle \omega| \, Y \widehat{\Psi}_2 \otimes \widehat{\Psi}_1 
	= - (-1)^{\widehat{\Psi}_1 \widehat{\Psi}_2 } \langle \widehat{\omega} | \, \widehat{\Psi}_2 \otimes \widehat{\Psi}_1 \, ,
\end{align}
following from~\eqref{eq:OGsymp} and~\eqref{eq:AntiCyc} for any $\widehat{\Psi}_i \in \widehat{\mathcal{H}}$. More importantly, it is non-degenerate~\emph{only} on $\widehat{\mathcal{H}}$
\begin{align}
	\forall \, \widehat{\Psi}_1 \in \widehat{\mathcal{H}} \quad \quad
	\langle \widehat{\omega} | \, \widehat{\Psi}_1 \otimes  \widehat{\Psi}_2 
	=\langle \omega | \widehat{\Psi}_1 \otimes Y \widehat{\Psi}_2= 0
	\implies  Y \widehat{\Psi}_2 = 0 
	\implies X Y \widehat{\Psi}_2 = \widehat{\Psi}_2 = 0 \, ,
\end{align} 
following from the non-degeneracy of $\langle \omega| $~\eqref{eq:nD} and requiring the restriction~\eqref{eq:C}. Lastly, the coderivation $\boldsymbol{m_1}$ can be shown to be cyclic with respect to $\langle \widehat{\omega}|$
\begin{align}
	\langle \widehat{\omega} | \, \widehat{\pi_2} \, \boldsymbol{m_1}
	&= \langle \omega | \, ( Y m_1 \otimes \id
	+  Y \otimes m_1) \nonumber \,= \langle \omega | \, (m_1 \otimes  Y 
	+ Y \otimes m_1 X Y )\nonumber  \,  \\
	&= \langle \omega | \, (- \id \otimes  m_1 Y 
	+  Y  \otimes X m_1 Y  ) \nonumber 
	= \langle \omega | \, (- XY \otimes  m_1 Y 
	+  XY  \otimes m_1 Y  ) = 0 \, ,
\end{align}
using the fact that it acts on $\widehat{\mathcal{H}}^{\otimes 2}$, so that $\id \to XY$, and by the equations~\eqref{eq:nD1} and~\eqref{eq:AntiCyc}. Not only $\boldsymbol{m_1}$ is cyclic, but all coderivations $\widehat{\boldsymbol{m_n}}$ have this property
\begin{align}
	\langle \omega | \, \widehat{\pi_2} \, \widehat{\boldsymbol{m_n}} = 
	\langle \omega | ( \widehat{m_n} \otimes \id + \id \otimes \widehat{m_n})  =0  \, .
\end{align}
This feature follows from the way the interactions in a particular theory is built in the presence of the constraint~\eqref{eq:C} on the states.

Finally, the action for the theory associated with the constrained cyclic $A_\infty$ algebra is given by
\begin{align} \label{eq:S}
	\widehat{S}[\widehat{\Psi}] &= \int_{0}^1 dt \, \,
	\langle \widehat{\omega} | \left[ \widehat{\pi_1} \, { \boldsymbol{d} \over \boldsymbol{dt}} \, {1 \over 1 - \widehat{\Psi}(t)} \otimes \widehat{\pi_1} \, \boldsymbol{m} \, {1 \over 1 - \widehat{\Psi}(t)} \right] 
	\nonumber \\
	&= {1 \over 2} \, \widehat{\omega}\left(\widehat{\Psi}, m_1 \widehat{\Psi} \right) 
	+ {1 \over 3}\, \widehat{\omega}\left(\widehat{\Psi}, \widehat{m_2}( \widehat{\Psi},\widehat{\Psi}) \right)
	+ {1 \over 4}\, \widehat{\omega}\left(\widehat{\Psi}, \widehat{m_3}( \widehat{\Psi}, \widehat{\Psi}, \widehat{\Psi})  \right) + \cdots
	\, ,
\end{align}
for which we define~\emph{the group-like element}
\begin{align}
	{1 \over 1 - \Psi} \equiv \sum_{n=0}^\infty \Psi^{\otimes n}
	= 1 + \Psi + \Psi \otimes \Psi + \cdots \, ,
\end{align}
and $\widehat{\Psi}(t)$ is a smooth function on $T \widehat{\mathcal{H}}$ that interpolates an element from  $t=0$ to $t=1$ assuming $\widehat{\Psi}(t=0) = 0$ and $\widehat{\Psi}(t=1) = \widehat{\Psi}$ is the (degree-even, grade-zero) dynamical field. Here 
\begin{align} \label{eq:partial}
	{ \boldsymbol{d} \over \boldsymbol{dt}} \equiv \sum_{n=1}^\infty \sum_{m=0}^{n-1} \id^{\otimes m} \otimes {d \over d t} \otimes \id^{\otimes (n-m-1)}
	=  {d \over d t} +  {d \over d t} \otimes \id + \id \otimes  {d \over d t}  + \cdots \, ,
\end{align}
is the degree-even coderivation associated with the derivative with respect to the parameter $t$. The equation of motion resulting from varying $\widehat{S}$ is 
\begin{align} \label{eq:eom0}
	0 = \widehat{\pi_1} \, \boldsymbol{m} \, {1 \over 1 - \widehat{\Psi}} 
	&= m_1 \widehat{\Psi}+ \widehat{m_2} \left(\widehat{\Psi}, \widehat{\Psi}\right) +
	\widehat{m_3} \left(\widehat{\Psi},\widehat{\Psi},\widehat{\Psi}\right) + \cdots  \, , 
\end{align}
while the infinitesimal gauge transformation is 
\begin{align} \label{eq:gauge}
	\delta  \widehat{\Psi}  = \widehat{\pi_1} \,  \boldsymbol{m} \, 
	{1 \over 1 - \widehat{\Psi}} \otimes \widehat{\Lambda} \otimes {1 \over 1 - \widehat{\Psi}}
	& = m_1 \widehat{\Lambda} +  \widehat{m_2} \left(\widehat{\Lambda} ,  \widehat{\Psi}\right) 
+  \widehat{m_2} \left(\widehat{\Psi}, \widehat{\Lambda} \right) + \cdots \, .
\end{align}
Here the field $\widehat{\Lambda}  \in \widehat{\mathcal{H}}$ is a (degree-odd, grade-zero) gauge parameter.

\section{Cyclic $A_\infty$ algebra without a constraint} \label{sec:noconstraint}

In this section we describe a cyclic homotopy algebra underlying the Sen's formalism. Recall that the action in Sen's formalism is given by~\cite{Sen:2015hha}
\begin{align} \label{eq:SenFree}
	S [\Psi, \widetilde{\Psi}]
	= -{1 \over 2} \omega(\widetilde{\Psi},  \mathcal{G} m_1 \widetilde{\Psi}) 
	+\omega(\widetilde{\Psi}, m_1 \Psi) + S_{int}[\Psi] \, ,
\end{align}
where $\Psi \in \mathcal{H}$ is the original field, now without any constraint, while (degree-even, grade-(-1)) $\widetilde{\Psi} \in \mathcal{H}$ is a spurious field. The worldsheet anomaly mentioned in previous section requires $\widetilde{\Psi}$ to have one less grade than $\Psi$. Here $S_{int}[\Psi] $ is the interaction term that only depends on the field $\Psi$. 

In the action~\eqref{eq:SenFree} $\mathcal{G}$ is a degree-even, anti-cyclic operator that raises the grade by one. It is assumed to commute with $m_1$, $[m_1, \mathcal{G}] = 0$. We can choose $\mathcal{G} = X$ of previous section, however this is not strictly required as long as the interactions are accounted correctly. For example it is more convenient to choose $\mathcal{G}$ as the zero mode of the picture changing operator $\mathcal{X} = [Q_B, \xi]$ for the Ramond sector of superstrings~\cite{Sen:2015hha}, which is different from~\eqref{eq:XY} that appears in the constraint~\cite{Erler:2016ybs}. We consider a generic $\mathcal{G}$ except for subsection~\ref{sec:hom}.

After varying~\eqref{eq:SenFree} we find
\begin{align} \label{eq:eom1}
	-\mathcal{G} m_1 \widetilde{\Psi} + m_1 \Psi   = 0 \, , \quad \quad
	m_1 \widetilde{\Psi} + J[\Psi] = 0 \, ,
\end{align}
where $J[\Psi]$ is the term results from varying $S_{int}[\Psi]$
\begin{align}
	\delta S_{int} = \omega(\delta \Psi, J[\Psi]) \, .
\end{align}
Acting on the second equation with $\mathcal{G}$ and adding to the first one in~\eqref{eq:SenFree} we obtain
\begin{align} \label{eq:int}
	m_1 \Psi + \mathcal{G} J[\Psi] = 0 \, .
\end{align}
Compare this with~\eqref{eq:eom0}: there was no need to impose a constraint on the field $\Psi$. This is due to the presence of $\widetilde{\Psi}$---it was impossible to write a free action just using the objects $\langle \omega |, \Psi, m_1$ while saturating the zero mode insertions required by the anomaly. However we pay the price of introducing a decoupled~\emph{free} field to our considerations through the first equation of motion~\eqref{eq:eom1}. The combination $\mathcal{G} \widetilde{\Psi} - \Psi$ can always be shifted by an unrelated field $\widetilde{\Sigma} \in \mathcal{H}$ that satisfies $m_1 \widetilde{\Sigma}   = 0 $.

Now define the space
\begin{align} \label{eq:Phi}
	\Phi = \begin{bmatrix}
		\Psi \\  \widetilde{\Psi}
	\end{bmatrix}
	\in
	\mathcal{K} \equiv \bigoplus_{n \in \mathbb{Z}}  \, ( \mathcal{H}_{n} \oplus \mathcal{H}_{n-1})\, ,
\end{align}
which makes the field $\Phi$ to have a uniform degree, i.e.
\begin{align}
	(-1)^{\Phi} = (-1)^{\Psi} = (-1)^{\widetilde{\Psi}} \, .
\end{align}
The subscript on $\mathcal{H}_n$ refers to the grade. It is convenient to define the degree-zero, grade-zero projections $P, \widetilde{P} : \mathcal{K} \to \mathcal{H}$
\begin{align} \label{eq:proj}
	P \Phi \equiv P \begin{bmatrix}
		\Psi \\ \widetilde{\Psi}
	\end{bmatrix} = \Psi \, , \quad \quad
	\widetilde{P} \Phi \equiv \widetilde{P} 
	\begin{bmatrix}
		\Psi \\ \widetilde{\Psi}
	\end{bmatrix} = \widetilde{\Psi} 
	\, .
\end{align}

The action~\eqref{eq:SenFree} can be cast to the $A_\infty$ language by introducing a new symplectic form
\begin{align}  \label{eq:simp}
	\langle \Omega | : \mathcal{K}^{\otimes 2}  \to \mathbb{C}^{1|1} \, , \quad \quad
	\langle \Omega | \equiv \langle \omega | \left[
		P\otimes   \widetilde{P} 
		+  \widetilde{P} \otimes P 
		-\widetilde{P} \otimes \mathcal{G} \widetilde{P}  \right] \, .
\end{align}
This form indeed has the expected properties. It is clearly bilinear, as well as it is anti-symmetric and the differential $M_1$ on $\mathcal{K}$ is cyclic
\begin{subequations} \label{eq:B}
\begin{align}
	&\langle \Omega | \,  \Phi_1 \otimes \Phi_2  = - (-1)^{\Phi_1 \Phi_2} \,\langle \Omega | \,  \Phi_2 \otimes \Phi_1 \, , \\ 
	&\langle \Omega | \left( M_1 \otimes \id + \id \otimes M_1 \right) = 0 \, ,
	\quad \quad 
		M_1 \Phi \equiv \begin{bmatrix}
			m_1 & 0 \\ 0 & m_1
		\end{bmatrix}
		\begin{bmatrix}
			\Psi \\ \widetilde{\Psi}
		\end{bmatrix}
		= \begin{bmatrix}
		m_1 \Psi \\  m_1 \widetilde{\Psi}
	\end{bmatrix} \, ,
\end{align}
\end{subequations}
as one can easily check. The identity $\id$ is the identity operator on $\mathcal{K}$ above. It will be apparent which identity we consider from the context. This symplectic form $\langle \Omega | $ is non-degenerate as well
\begin{align}
	\forall \, \Phi_1 \in \mathcal{K} \,  \quad \quad
	\langle \Omega | \,  \Phi_1 \otimes \Phi_2 &=
	\langle \omega | \left[ \Psi_1 \otimes \widetilde{\Psi}_2
	+ \widetilde{\Psi}_1 \otimes \Psi_2 
	- \widetilde{\Psi}_1 \otimes \mathcal{G} \widetilde{\Psi}_2 \right] = 0
	\nonumber \\
	&\implies \Psi_2 = \widetilde{\Psi}_2 = 0 \implies \Phi_2 = 0 \, ,
\end{align}
given that the terms proportional to $\Psi_1$ and $\widetilde{\Psi}_1$ can be individually set to zero. 

We can rewrite the action~\eqref{eq:eom1} using $\langle \Omega |$ and $\Phi$ in the $A_\infty$ language and find the equations of motion expected from the free theory, i.e.~\eqref{eq:eom1} when $J = 0$. However the case with interactions is far more interesting. These are described by the degree-odd, multi-linear maps $m_n: \mathcal{H}^{\otimes n} \to \mathcal{H}$ for $n \geq 2$. They are taken to be cyclic with respect to the form $\langle \omega |$ on $\mathcal{H}$
\begin{align} \label{eq:mcyc}
	\langle \omega | \left[ m_n \otimes \id + \id \otimes m_n \right] = 0 \, .
\end{align}
and satisfy a defining identity that we will come in a moment. It is useful to introduce a related set of degree-odd multi-linear maps
\begin{align} \label{eq:Gm}
	\widecheck{m_n}: \mathcal{H}^{\otimes n} \to \mathcal{H} \, , \quad \quad
	\widecheck{m_n} = \mathcal{G} m_n \, , 
\end{align}
and uplift them to the coalgebra $T \widehat{\mathcal{H}}$ as degree-odd coderivations
\begin{align}
	\boldsymbol{\widecheck{m_n}} = \boldsymbol{\pi} [ \boldsymbol{\mathcal{G}} \boldsymbol{m_n} ] \, ,
\end{align} 
in order to facilitate the discussion of this identity. Above $\boldsymbol{\pi}$ is the formal multi-linear operation
\begin{align} \label{eq:1}
	\boldsymbol{\pi}[ \cdots \otimes \mathcal{G} \otimes \cdots] = \cdots \otimes \id \otimes \cdots \, ,
\end{align}
that is necessary to turn $\boldsymbol{\widecheck{m_n}}$ into a coderivation\footnote{These types of formal objects that replace the operators inside the expressions to obtain the desired cooperations have been introduced recently in the context of the stubbed SFTs~\cite{Schnabl:2023dbv,Schnabl:2024fdx,Erbin:2023hcs,Maccaferri:2024puc}.} and $\boldsymbol{\mathcal G}$ is the cohomomorphism constructed using $\mathcal{G}$ like in~\eqref{eq:Coho}. We have $[\boldsymbol{m_1}, \boldsymbol{\mathcal{G}}] = \boldsymbol{0}$ similar to~\eqref{eq:like}. We take
\begin{align} \label{eq:2}
	\boldsymbol{\widecheck{\delta m}} = \boldsymbol{\pi}[ \boldsymbol{\mathcal{G}} \boldsymbol{\delta m}] \, , \quad \quad
	\boldsymbol{\delta m } 
	\equiv \sum_{n=2}^\infty \boldsymbol{m_n} \equiv \boldsymbol{m_2} + \boldsymbol{m_3} + \cdots \, ,
	\quad
\end{align}
and 
\begin{align} \label{eq:mcheck}
	\boldsymbol{\widecheck m} \equiv \boldsymbol{m_1} + \boldsymbol{\widecheck{\delta m}}
	= \boldsymbol{m_1} + \boldsymbol{\pi}[ \boldsymbol{\mathcal{G}} \boldsymbol{\delta m}]  \, .
\end{align}
Note that the output grades of $m_n$ are one lower than those of $\widecheck{m_n}$ by the presence of $\mathcal{G}$~\eqref{eq:Gm}.

The aforementioned identity satisfied by the multi-linear maps $m_n$ is then encoded by
\begin{align} \label{eq:A2}
	\pi_1 [ \boldsymbol{m_1} \, \boldsymbol{\delta m}  
	+\boldsymbol{\delta m}  \, \boldsymbol{\widecheck{m}} ]
	&=
	\pi_1 [ \boldsymbol{m_1} \, \boldsymbol{\delta m} 
	+ \boldsymbol{\delta m} \, \boldsymbol{m_1}
	+\boldsymbol{\delta m} \, \boldsymbol{ \widecheck{\delta m} } ]
	\nonumber \\
	&=
	\pi_1[ \boldsymbol{m_1} \, \boldsymbol{\delta m} 
	+ \boldsymbol{\delta m} \, \boldsymbol{m_1}
	+\boldsymbol{\delta m} \, \boldsymbol{\pi} [ \boldsymbol{\mathcal{G}} \boldsymbol{\delta m} ] ]=  0\, .
\end{align}
The gauge invariance in Sen's formalism is a consequence of this relation~\cite{Erler:2016ybs}. Here $\pi_n$ is the projection on $T\mathcal{H}$, that is $\pi_n T\mathcal{H} = \mathcal{H}^{\otimes n}$. We highlight that constructing $m_n$ with the properties~\eqref{eq:mcyc} and~\eqref{eq:A2} may not be straightforward a priori and require an input from the theory under consideration~\cite{Erler:2016ybs,Okawa:2022mos}. We assume this can be done for our purposes.

An important corollary to~\eqref{eq:A2} shows that the coderivation $\boldsymbol{\widecheck{m}}$~\eqref{eq:mcheck} itself is nilpotent
\begin{align}\label{eq:A3}
	0 &= \mathcal{G} \pi_1 [ \boldsymbol{m_1} \, \boldsymbol{\delta m} 
	+ \boldsymbol{\delta m} \, \boldsymbol{m_1}
	+\boldsymbol{\delta m} \, \boldsymbol{ \widecheck{\delta m} } ]
	\nonumber \\
	&= \pi_1 [ \boldsymbol{m_1} \, \boldsymbol{\widecheck{\delta m}} 
	+ \boldsymbol{\widecheck{\delta m}} \, \boldsymbol{m_1}
	+\boldsymbol{\widecheck{\delta m}} \, \boldsymbol{ \widecheck{\delta m} } ]
	= \pi_1 [\boldsymbol{\widecheck{m}}, \boldsymbol{\widecheck{m}}]
	\quad \implies \quad
	\boldsymbol{\widecheck{m}}^2 = \boldsymbol{0} \, ,
\end{align}
where we used $[m_1, \mathcal{G}] = 0$, $\boldsymbol{m_1}^2 = \boldsymbol{0}$, and~\eqref{eq:mcheck}. Importantly, however, the coderivation $\boldsymbol{\widecheck{m}}$ isn't cyclic. The construction below is going to bypass this issue as we shall see.

At this point it is beneficial to make a remark on the relation between $\boldsymbol{\widecheck{m}}$ and $\boldsymbol{m}$ of the constrained theory~\eqref{eq:m} in previous section. They are supposed to be the same upon choosing $\mathcal{G} = X$, which occurs when the maps $m_n$ and $\widehat{m_n}$ are related by
\begin{align} \label{eq:pro}
	\widehat{m_n} = X m_n \, , \quad \quad
	n \geq 2 \, .
\end{align}
The image of $X m_n$ always belongs to $\widehat{\mathcal{H}}$ due to the identity $XYX = X$ so the maps $\widehat{m_n}$ remain well-defined on $\widehat{\mathcal{H}}$. The nilpotency~\eqref{eq:A3} is automatic in this case by~\eqref{eq:m0}. The constructions along these lines for the maps satisfying this type of relations for the open superstrings can be found in~\cite{Erler:2016ybs} and for the closed bosonic strings without level-matching can be found in~\cite{Okawa:2022mos}.

Now we are ready to define the degree-odd multi-linear maps $M_n : \mathcal{K}^{\otimes n} \to \mathcal{K}$ for $n\geq2$
\begin{align} \label{eq:expilicit}
	M_n(\Phi_1, \cdots, \Phi_n) \equiv \begin{bmatrix}
		\mathcal{G} m_n (P \Phi_1, \cdots, P \Phi_n) \\
		m_n (P \Phi_1, \cdots, P \Phi_n)
	\end{bmatrix} 
	 = \begin{bmatrix}
		\widecheck{m_n} (P \Phi_1, \cdots, P \Phi_n) \\
		m_n (P \Phi_1, \cdots, P \Phi_n)
	\end{bmatrix} \, .
\end{align}
Note that they have the correct grades at each row. These can be further expressed on the coalgebra $T \mathcal{K}$ as degree-odd coderivations
\begin{align}
	\boldsymbol{M_n} = 
		\boldsymbol{\Pi}[ \,
		\boldsymbol{\iota}  \boldsymbol{\mathcal{G}} \boldsymbol{m_n}  \boldsymbol{P} +
		\boldsymbol{\widetilde{\iota}}  \boldsymbol{m_n}  \boldsymbol{P} \, ] 
		=
		\boldsymbol{\Pi}[ \,
		\boldsymbol{\iota} \boldsymbol{\widecheck{m_n}}  \boldsymbol{P} +
		\boldsymbol{\widetilde{\iota}}  \boldsymbol{m_n}  \boldsymbol{P} \, ] \, ,
\end{align}
where $\iota, \widetilde{\iota} : \mathcal{H} \to \mathcal{K}$ are the canonical inclusion maps to the first and second factors of the space $\mathcal{K}$~\eqref{eq:Phi} respectively
\begin{align} \label{eq:inc}
	\iota  \, \Psi  \equiv \begin{bmatrix}
		\Psi \\ 0 
	\end{bmatrix} \, , \quad \quad
	\widetilde{\iota} \,  \Psi  \equiv \begin{bmatrix}
		0 \\ \Psi
	\end{bmatrix} \, ,
\end{align}
for $\Psi \in \mathcal{H}$ and their associated cohomomorphisms defined similar to~\eqref{eq:Coho}. Clearly 
\begin{align} \label{eq:Pi}
	P \iota = \widetilde{P} \widetilde{\iota} = 1 \, , \quad \quad
	P \widetilde{\iota} = \widetilde{P} \iota = 0 
	\quad \implies \quad
	\boldsymbol{P} \boldsymbol{\iota} = \boldsymbol{\widetilde{P}} \boldsymbol{\widetilde{\iota}} = \boldsymbol{1} \, , \quad \quad
	\boldsymbol{P} \boldsymbol{\widetilde{\iota}} = \boldsymbol{\widetilde{P}}\boldsymbol{\iota}  = \Pi_0
	\, .
\end{align}
 Here $\Pi_n$ is the projection on $T\mathcal{K}$, that is $\Pi_n T\mathcal{K} = \mathcal{K}^{\otimes n}$. The operation $\boldsymbol{\Pi}$ is given by
\begin{align}
	\boldsymbol{\Pi} [ \cdots \otimes \iota \mathcal{G} P \otimes \cdots] =  
	\boldsymbol{\Pi} [ \cdots \otimes \iota P \otimes \cdots] =
	\boldsymbol{\Pi} [ \cdots \otimes \widetilde{\iota} P \otimes \cdots] =  \cdots \otimes \id \otimes \cdots \, ,
\end{align}
which makes $\boldsymbol{M_n} $ coderivation. We collect the coderivations
\begin{align} \label{eq:M0}
	\boldsymbol{M}
	=\boldsymbol{M_1} + \sum_{n=2}^\infty \boldsymbol{M_n}
	=   \boldsymbol{M_1}+  \boldsymbol{M_2} + \cdots 
	= \boldsymbol{M_1} + \boldsymbol{\delta M}
	\, , 
\end{align}
to write interactions compactly as
\begin{align} \label{eq:M}
	\boldsymbol{\delta M} = \boldsymbol{\Pi}[ \, \boldsymbol{\iota} \boldsymbol{\mathcal{G}}  \boldsymbol{\delta m }  \boldsymbol{P} +\boldsymbol{\widetilde{\iota}}  \boldsymbol{\delta m }  \boldsymbol{P} \, ] 
	=\boldsymbol{\Pi}[ \, \boldsymbol{\iota} \boldsymbol{\widecheck{\delta m}}  \boldsymbol{P} + \boldsymbol{\widetilde{\iota}}  \boldsymbol{\delta m }  \boldsymbol{P} \, ] 
	\, ,
\end{align}
where $\boldsymbol{M_1}$ is constructed using $M_1$ similar to~\eqref{eq:partial}. It is clearly odd and nilpotent.

Let us demonstrate two facts now: the odd coderivation $\boldsymbol{M}$~\eqref{eq:M0} is nilpotent and it is cyclic with respect to the symplectic form $\langle \Omega |$~\eqref{eq:simp}. Begin with showing the former. First of all
\begin{subequations} \label{eq:all}
\begin{align}
	\boldsymbol{M_1} \boldsymbol{\delta M}  + \boldsymbol{\delta M} \boldsymbol{M_1} 
	&= \boldsymbol{M_1} \boldsymbol{\Pi}[ \, \boldsymbol{\iota} \boldsymbol{\widecheck{\delta m}}  \boldsymbol{P} +\boldsymbol{\widetilde{\iota}} \boldsymbol{\delta m }  \boldsymbol{P} \, ]  +  \boldsymbol{\Pi}[ \, \boldsymbol{\iota} \boldsymbol{\widecheck{\delta m}}  \boldsymbol{P} +\boldsymbol{\widetilde{\iota}} \boldsymbol{\delta m }  \boldsymbol{P} \, ] \boldsymbol{M_1}
	\nonumber \\
	&= \boldsymbol{\Pi}[ \, \boldsymbol{\iota} (\boldsymbol{m_1} \, \boldsymbol{\widecheck{\delta m} }+ \boldsymbol{\widecheck{\delta m}} \, \boldsymbol{m_1}    ) \boldsymbol{P} 
	+ \boldsymbol{\widetilde{\iota}} (\boldsymbol{m_1} \, \boldsymbol{\delta m } + \boldsymbol{\delta m }  \, \boldsymbol{m_1} )\boldsymbol{P} \, ]
	\, ,
\end{align}
where  $\boldsymbol{M_1}$ is moved inside the operation $\boldsymbol{\Pi}$ and commuted with the projections and inclusions, which resulted $\boldsymbol{m_1} $ inside $\boldsymbol{\Pi}$ as shown above, see~\eqref{eq:proj},~\eqref{eq:B}, and~\eqref{eq:inc}. We also evaluate
\begin{align}
	\boldsymbol{\delta M}^2 =\boldsymbol{\Pi}[ \, \boldsymbol{\iota}  \boldsymbol{\widecheck{\delta m}}  \boldsymbol{P}
	+\boldsymbol{\widetilde{\iota}} \boldsymbol{\delta m }  \boldsymbol{P} \, ]
	\boldsymbol{\Pi}[ \, 
	 \boldsymbol{\iota} \boldsymbol{\widecheck{\delta m}}  \boldsymbol{P}
	+ \boldsymbol{\widetilde{\iota}} \boldsymbol{\delta m }  \boldsymbol{P} 
	\, ] =
	\boldsymbol{\Pi} [ \boldsymbol{\iota}  \boldsymbol{\widecheck{\delta m}}^2  \boldsymbol{P} +
	\boldsymbol{\widetilde{\iota}} \boldsymbol{\delta m } \, \boldsymbol{\widecheck{\delta m} } \boldsymbol{P} ] \, ,
\end{align}
\end{subequations}
where we used~\eqref{eq:Pi}. The ``cross terms'' in both expressions are canceled by the anti-commutation. Combining~\eqref{eq:all} and using the relations~\eqref{eq:A2}-\eqref{eq:A3} we get
\begin{align}
	\Pi_1 [ \boldsymbol{M_1} \boldsymbol{\delta M}  + \boldsymbol{\delta M} \boldsymbol{M_1}  + \boldsymbol{\delta M}^2 ] = 0\quad &\implies \quad
	\Pi_1 [ \boldsymbol{M}, \boldsymbol{M}]  = 0 \quad 
	\nonumber \\
	&\implies \quad
	\boldsymbol{M}^2 = \boldsymbol{0}
	\, .
\end{align}
In the second step we have included $\boldsymbol{M_1}^2 =\boldsymbol{0}$ and in the last step we have used $\boldsymbol{M}$ is a coderivation from the construction. Indeed, $\boldsymbol{M}$ is an odd~\emph{nilpotent} coderivation on the tensor coalgebra $T \mathcal{K}$.

Next we check the cyclicity of $\boldsymbol{M_n}$ with respect to the symplectic form $\langle \Omega |$.  This is easier to accomplish with the form given in~\eqref{eq:expilicit}. So focus on
\begin{subequations} \label{eq:all2}
\begin{align}
	\langle \Omega | \,  (M_n \otimes \id) \, (\Phi_1, \cdots, \Phi_{n+1}) 
	&= \langle \Omega | \,  M_n(\Phi_1, \cdots, \Phi_n) \otimes \Phi_{n+1}
	= \langle \Omega | \begin{bmatrix}
		\mathcal{G} m_n (P \Phi_1, \cdots, P \Phi_n) \nonumber \\
		m_n (P \Phi_1, \cdots, P \Phi_n)
	\end{bmatrix} \otimes \Phi_{n+1} \\
	&\hspace{-5em}= \langle \omega | \left[ \mathcal{G}  m_n (\Psi_1, \cdots, \Psi_n)  \otimes \widetilde{\Psi}_{n+1}
	+m_n (\Psi_1, \cdots, \Psi_n) \otimes (\Psi_{n+1} -\mathcal{G}  \widetilde{\Psi}_{n+1}) \right] \nonumber \\
	&\hspace{-5em}= \langle \omega | \, m_n (\Psi_1, \cdots, \Psi_n) \otimes \Psi_{n+1} 
	=\langle \omega | \,  (m_n \otimes \id) \, (\Psi_1, \cdots, \Psi_{n+1}) 
	\, ,
\end{align}
where we have used~\eqref{eq:simp} and the fact that $\mathcal{G} $ is anti-cyclic. Analogous arguments establish
\begin{align}
	\langle \Omega | \,  (\id \otimes M_n) \, (\Phi_1, \cdots, \Phi_{n+1}) 
	&=
	 (-1)^{\Phi_1} \langle \Omega | \, \Phi_1 \otimes M_n ( \Phi_2 \cdots, \Phi_{n+1}) 
	 \nonumber \\
	&=  (-1)^{\Psi_1}  \langle \omega | \, \Psi_1  \otimes m_n ( \Psi_2 \cdots, \Psi_{n+1}) 
	\nonumber \\
	&=\langle \omega | \,  (\id \otimes m_n) \, (\Psi_1, \cdots, \Psi_{n+1}) \, .
\end{align}
\end{subequations}
Combining~\eqref{eq:all2} together shows $\boldsymbol{M_n} $  is indeed cyclic with respect to the symplectic form $\langle \Omega | $
\begin{align} \label{eq:MC}
	\langle \Omega | \left( M_n \otimes \id + \id \otimes M_n \right) =
	\langle \Omega | \, \Pi_2 \boldsymbol{M_n} =  0 
	\quad \implies \quad \langle \Omega | \, \Pi_2 \boldsymbol{M} =  0 \, .
\end{align}
from~\eqref{eq:B} and~\eqref{eq:mcyc}.

As promised a cyclic $A_\infty$ algebra is obtained for Sen's formalism, see~\eqref{eq:simp} and~\eqref{eq:M}. The associated action is
\begin{align}
		S[\Phi] &= \int_{0}^1 dt \, \,
		\langle \Omega | \left[ \Pi_1 { \boldsymbol{d} \over \boldsymbol{dt}} {1 \over 1 - \Phi(t)} \otimes \Pi_1 \boldsymbol{M} {1 \over 1 - \Phi(t)} \right] 
		\nonumber \\
		&= {1 \over 2} \Omega(\Phi,  M_1 \Phi) 
		+ {1\over 3} \Omega(\Phi, M_2 (\Phi, \Phi))+\cdots
		\nonumber \\
		&= -{1 \over 2} \omega(\widetilde{\Psi},  \mathcal{G} m_1 \widetilde{\Psi}) 
		+\omega(\widetilde{\Psi}, m_1 \Psi) + {1\over 3} \omega(\Psi, m_2 (\Psi, \Psi))+\cdots
		\, ,
\end{align}
which is the same as~\eqref{eq:SenFree}. The dynamical field $\Phi$ is degree-even and taken to belong $\mathcal{H}_{0} \oplus \mathcal{H}_{-1}$. It is also interesting to report the equation of motion
\begin{align} \label{eq:eomlast}
	0 = \Pi_1 \boldsymbol{M} {1 \over 1 - \Phi} 
	&=  \sum_{n=1}^\infty M_n (\Phi^{\otimes n}) = \begin{bmatrix}
		m_1 \Psi + \mathcal{G} m_2(\Psi, \Psi) + \mathcal{G}  m_3 (\Psi, \Psi, \Psi) + \cdots \\
		m_1 \widetilde{\Psi} + m_2 (\Psi, \Psi ) + m_3 (\Psi, \Psi, \Psi) +\cdots
	\end{bmatrix} \, ,
\end{align}
which is precisely what we would have obtained upon choosing
\begin{align}
	J[\Psi] = \sum_{n=2}^\infty m_n(\Psi^{\otimes n})
	= m_2(\Psi, \Psi) + m_3(\Psi, \Psi, \Psi) + \cdots \, ,
\end{align}
in~\eqref{eq:eom1} and~\eqref{eq:int}. The gauge transformations can be given like in~\eqref{eq:gauge} as well
\begin{align} \label{eq:gauge0}
	\delta  \Phi  &= \Pi_1 \,  \boldsymbol{M} \, 
	{1 \over 1 - \Phi} \otimes \Omega \otimes {1 \over 1 - \Phi}
	\nonumber \\
	& = M_1 \, \Omega +  M_2 \left(\Omega ,  \Phi\right) 
	+  M_2 \left(\Phi, \Omega \right) + \cdots 
	= \begin{bmatrix}
		m_1 \Lambda + \mathcal{G}  m_2 \left(\Lambda ,  \Psi \right) 
		+  \mathcal{G}  m_2 \left(\Psi, \Lambda \right) + \cdots 
		\\ m_1 \widetilde{\Lambda} + m_2 \left(\Lambda ,  \Psi \right) +  m_2 \left(\Psi, \Lambda \right)  + \cdots
	\end{bmatrix}\, ,
\end{align}
where the field $\Omega$ is a degree-odd gauge parameter given by
\begin{align}
	\Omega = \begin{bmatrix}
		\Lambda \\ \widetilde{\Lambda} 
	\end{bmatrix} \in \mathcal{H}_{0} \oplus \mathcal{H}_{-1} \, .
\end{align}
This is indeed the expected gauge transformation for the action~\eqref{eq:SenFree}. Note that it contains the extra gauge symmetry with the parameter $\widetilde{\Lambda}$.

\subsection{Homotopy transfer to the constrained algebra} \label{sec:hom}

As an application we generate the constrained algebra from the unconstrained counterpart in this subsection. In order to do that we need to take $\mathcal{G}  = X$ and~\emph{impose} the constraints
\begin{align} \label{eq:cons}
	\widetilde{\Psi}  - Y \Psi = \widetilde{\Sigma} \, , \quad \quad
	X \widetilde{\Psi} - \Psi = X \widetilde{\Sigma}  \, , \quad \quad
	m_1 \widetilde{\Sigma} = 0 \, ,
\end{align}
between the factors of $\mathcal{K}$~\eqref{eq:Phi}. As we shall see shortly, the string field $\widetilde \Sigma$ is going to be associated with the Sen's decoupled free modes. The equation~\eqref{eq:cons} immediately requires
\begin{align} \label{eq:cons1}
	XY | \Psi \rangle =  | \Psi \rangle  \, ,
\end{align}
for the fields in the first row of $\Phi$. This is the avatar of the original restriction~\eqref{eq:C} in $\mathcal{K}$. Observe that all dependence on the $m_1$-closed fields $\widetilde{\Sigma}$ is irrelevant for this constraint.

The restrictions~\eqref{eq:cons} should be imposed on top of the action in the vein similar to self-dual constraints on the $p$-form fields. We highlight that the right-hand sides of them contain an arbitrary $m_1$-closed (or~\emph{free}) fields in general. Directly setting  $\widetilde{\Psi} - Y \Psi$ or $\Psi - X \widetilde{\Psi} $ to zero would lead to an inconsistency due to their combination can contain $m_1$-closed, but~\emph{not} necessarily $m_1$-exact, fields. These modes can't be simply set to zero. Furthermore~\eqref{eq:cons} can't be thought as a gauge-fixing condition as explained in~\cite{Erbin:2020eyc}. These are genuine constraints.

An insightful reader may already realize the field $ \widetilde{\Sigma} $ here is associated with the Sen's decoupled free field. This is indeed the case. In order to see this precisely identify the elements $\Phi \in \mathcal{K}$ constrained by~\eqref{eq:cons} through
\begin{align} \label{eq:ident}
	\Phi = \begin{bmatrix}
		\Psi \\ Y \Psi
	\end{bmatrix}
	\simeq \begin{bmatrix}
		\Psi \\ Y \Psi
	\end{bmatrix} +
	\begin{bmatrix}
		0 \\ \widetilde{\Sigma}
	\end{bmatrix} = \Phi + 
	\begin{bmatrix}
		0 \\ \widetilde{\Sigma}
	\end{bmatrix} 
	\, , \quad \quad
	m_1 \widetilde{\Sigma} = 0 
	\, .
\end{align}
This identification instructs to treat the fields $\Phi$ that differ by $ \widetilde{\Sigma}$ like above in the same way and the free field $ \widetilde{\Sigma}$ would decouple from our constructions for all intents and purposes as a result. This is the behavior expected from the Sen's free field. We denote the quotient defined by~\eqref{eq:ident} $\widehat{\mathcal{K}} = \mathcal{K}/ \simeq$.

It is highly crucial to emphasize the identification $\simeq$ above does~\emph{not} eliminate the decoupled free modes from the physical spectrum: that is, we are~\emph{not} integrating them out through homotopy transfer. We are simply treating the original interacting modes differing by Sen's spurious free modes equivalent by working on $\widehat{\mathcal{K}}$. Therefore the identification $\simeq$ should be understood as a mathematical statement of the decoupling of Sen's free modes from the rest of the theory, at least perturbatively. This makes sense: we shouldn't have any way of detecting Sen's spurious fields from the perspective of the constrained theory.

Now it is possible to construct the (degree-even, grade-zero) bijection $F$ between $\widehat{\mathcal{H}}$ and $\widehat{\mathcal{K}}$
\begin{align} \label{eq:F}
	F : \widehat{\mathcal{H}} \to\widehat{\mathcal{K}} \, , \quad \quad
	F \widehat{\Psi} =
	 \begin{bmatrix}
		1\\ Y  
	\end{bmatrix} \widehat{\Psi} = 
	 \begin{bmatrix}
		\widehat{\Psi} \\ Y  \widehat{\Psi}
	\end{bmatrix}  \, ,
\end{align}
whose inverse $F^{-1}$ is well-defined thanks to the quotient~\eqref{eq:ident}. We can lift  the bijection $F$ to the level of the tensor coalgebra as coisomorphism~$\boldsymbol{F}$ via~\eqref{eq:Coho} upon changing $X \to F$. Then the coderivation $\boldsymbol{M}$ of $T\widehat{\mathcal{K}}$ can be pushed to $ T\widehat{\mathcal{H}}$ by
\begin{align} \label{eq:FF}
	 \boldsymbol{\widehat{ m} }  = \boldsymbol{F^{-1}} \boldsymbol{M} \boldsymbol{F} \, ,
\end{align}
where we denoted the odd coderivation on $T\widehat{\mathcal{H}}$ by $\boldsymbol{\widehat{ m} }$. Note that $\boldsymbol{M}$ (and $\langle \Omega |$) trivially descends from $T\mathcal{K}$ to the quotient $T\widehat{\mathcal{K}}$: acting the vertices with the fields differing by~\eqref{eq:ident} would get the same answer because of the projector $P$ in the definition~\eqref{eq:expilicit}. Since the identification $\simeq$ does not correspond to mode integration as we stressed above, the products remain unaltered. In other words, the theory described by the products $\boldsymbol{\widehat{ m} }$ is not an~\emph{effective} theory, but simply a~\emph{constrained} one from the perspective of $TK$. The nilpotency of the coderivation is preserved under the map $F$
\begin{align}
	\boldsymbol{\widehat{ m} }^2 =  \boldsymbol{F^{-1}} \boldsymbol{M} \boldsymbol{F}  \boldsymbol{F^{-1}} \boldsymbol{M} \boldsymbol{F} = \boldsymbol{F^{-1}} \boldsymbol{M}^2 \boldsymbol{F}  = \boldsymbol{0} \, ,
\end{align}
which states the $A_\infty$ morphism between $\widehat{\mathcal{K}}$ and $\widehat{\mathcal{H}}$.

It shouldn't be too surprising to learn
\begin{align}
	\boldsymbol{\widehat{ m} }  = \boldsymbol{ m } =
	\boldsymbol{m_1} + \boldsymbol{ \widehat{m_2}} + \boldsymbol{ \widehat{m_3}} + \cdots  \, ,
\end{align}
after observing~\eqref{eq:expilicit} and~\eqref{eq:F}. Moreover the coderivation $\boldsymbol{\widehat{ m} } $ is cyclic with respect to the symplectic form
\begin{align}
	\langle \widehat{\Omega}|: \widehat{\mathcal{H}}^{\otimes 2}  \to \mathbb{C}^{1|1} \, , \quad \quad \langle \widehat{\Omega} | \, \widehat{\pi_2}  = \langle \Omega |\, \Pi_2 \boldsymbol{F} \, .
\end{align}
This is trivial to establish
\begin{align}
	\langle \widehat{\Omega} | \widehat{\pi_2} \, \boldsymbol{\widehat{m}} = 
	\langle \Omega| \Pi_2 \boldsymbol{F} \, \boldsymbol{\widehat{m}} =
	\langle \Omega| \Pi_2 \boldsymbol{M} \boldsymbol{F} = 0 \, ,
\end{align}
using~\eqref{eq:MC}~\eqref{eq:FF} ~\eqref{eq:F} and descending the projections $\Pi_n$ to $T \widehat{\mathcal{K}}$. In fact $\langle \widehat{\Omega} | \widehat{\pi_2} $ evaluates to
\begin{align}
	\langle \widehat{\Omega} | \, \widehat{\pi_2} =
	\langle \Omega | \, \Pi_2 \boldsymbol{F} &= 
	\langle \omega | \,\left[
	P\otimes   \widetilde{P} 
	+  \widetilde{P} \otimes P 
	-\widetilde{P} \otimes X \widetilde{P}  \right] \begin{bmatrix}
		1 \\ Y
	\end{bmatrix} 
	\otimes
	\begin{bmatrix}
		1 \\ Y
	\end{bmatrix} \, \widehat{\pi_2}
	\nonumber \\
	&= \langle \omega | \, (\id \otimes Y + Y \otimes \id - Y \otimes X Y) \, \widehat{\pi_2}
	= \langle \omega | \, (Y \otimes \id) \, \widehat{\pi}_2= \langle \widehat{\omega} |\, \widehat{\pi_2}
	\, ,
\end{align}
after using~\eqref{eq:AntiCyc},~\eqref{eq:simp}. This shows the form $\langle \widehat{\Omega} | $ is nothing other than the symplectic form of the constrained algebra~\eqref{eq:what} and we have indeed generated the constrained cyclic $A_\infty$ algebra from the unconstrained one. 

The action after the homotopy transfer is given by~\eqref{eq:S} with the additional (and expected) constraint~\eqref{eq:cons1} on the dynamical field $\widehat{\Psi}$. In order to see this concretely note that we have
\begin{align}
	\boldsymbol{F} \, {1 \over 1 - \widehat{\Psi}(t)}  = {1 \over 1-F(\widehat{\Psi}(t))} \, , 
\end{align}
for a smooth interpolation between $0 \leq t \leq 1$ with $\widehat{\Psi}(0) = 0$ and $\widehat{\Psi}(1) = \widehat{\Psi} \in \widehat{\mathcal{H}}$. Then imposing the constraints in~\eqref{eq:cons} on the form of the field $\Phi = \Phi (t)$ leads to
\begin{align}
	S[\Phi = F(\widehat{\Psi})] 
	&= \int_{0}^1 dt \, \,
	\langle \Omega | \left[ \Pi_1 { \boldsymbol{d} \over \boldsymbol{dt}} \boldsymbol{F} {1 \over 1 - \widehat{\Psi}(t)} \otimes \Pi_1 \boldsymbol{M} \boldsymbol{F} {1 \over 1 - \widehat{\Psi}(t)} \right] 
	\nonumber \\
	&= \int_{0}^1 dt \, \,
	\langle \Omega | \left[ \Pi_1 \boldsymbol{F} { \boldsymbol{d} \over \boldsymbol{dt}} {1 \over 1 - \widehat{\Psi}(t)} \otimes \Pi_1 \boldsymbol{F}\, \boldsymbol{\widehat{m}} {1 \over 1 - \widehat{\Psi}(t)} \right] 
	\nonumber \\
	&=\int_{0}^1 dt \, \,
	\langle \widehat{\Omega} | \left[ \widehat{\pi_1} { \boldsymbol{d} \over \boldsymbol{dt}} {1 \over 1 - \widehat{\Psi}(t)} \otimes \widehat{\pi_1}  \boldsymbol{m} {1 \over 1 - \widehat{\Psi}(t)} \right] = \widehat{S}[\widehat{\Psi}] \, ,
\end{align}
using various relations from above. This is the constrained action~\eqref{eq:S} and the field $\widehat{\Psi}$ satisfies the constraint~\eqref{eq:cons1} by construction. On top of this we also have a decoupled set of free modes and they become invisible from the perspective of the constrained theory.

It may be interesting to investigate different types of homotopy transfers, especially its possible relation to the Feynman graph interpretation of the Ramond sector interactions~\cite{Erler:2017onq}, within the context of the unconstrained algebra but we conclude our discussion here.

\section*{Acknowledgments}
The author thanks Manki Kim, Raji Mamade, Nicol\'as Vald\'es-Meller, and Barton Zwiebach for the assorted discussions on related topics and Harold Erbin for discussions and his comments on the early draft. We also would like to thank Ted Erler for his insightful comments and bringing the author's attention to~\cite{Erler:2017pgf}. This material is based upon work supported by the U.S. Department of Energy, Office of Science, Office of High Energy Physics of U.S. Department of Energy under grant Contract Number  DE-SC0012567.


\providecommand{\href}[2]{#2}\begingroup\raggedright\endgroup

\end{document}